\documentclass[aps,prd,showpacs,amssymb,superscriptaddress, notitlepage,floats,floatfix,nofootinbib, twocolumn]{revtex4-2}

\usepackage{graphicx}
\usepackage{dcolumn}
\usepackage{bm}
\usepackage{amsmath}
\usepackage{commath}
\usepackage{color}
\usepackage[dvipsnames]{xcolor}
\usepackage{hyperref}
\hypersetup{colorlinks=true, citecolor=MidnightBlue,
            linkcolor=CornflowerBlue, urlcolor=CornflowerBlue, linktocpage=true}
\usepackage{journals}
\usepackage{booktabs}
\usepackage{wasysym}
\usepackage[normalem]{ulem}
\usepackage{soul}

\definecolor{darkgreen}{rgb}{0.0, 0.6, 0.0}

\begin{document}

\title{Constraining the Nuclear Equation of State from Rotating Neutron Stars}

\author{Sebastian H. V\"olkel}
\email{sebastian.voelkel@sissa.it}

\affiliation{SISSA - Scuola Internazionale Superiore di Studi Avanzati, via Bonomea 265, 34136 Trieste, Italy and INFN Sezione di Trieste}
\affiliation{IFPU - Institute for Fundamental Physics of the Universe, Via Beirut 2, 34014 Trieste, Italy}

\author{Christian J. Kr\"uger}
\email{christian.krueger@tat.uni-tuebingen.de}

\affiliation{Theoretical Astrophysics, IAAT, University of T\"ubingen, 72076 T\"ubingen, Germany}

\date{\today}

\begin{abstract}
We demonstrate how observables of slowly rotating neutron stars can be used to constrain the nuclear equation of state. By building a Bayesian framework we demonstrate how combining different types of neutron star measurements, motivated by the upcoming multi-messenger era, provide informative posterior distributions of commonly used equation of state parametrizations. Exemplarily, we chose the Read \textit{et al.} four parameter model that is widely used to represent realistic equations of states in neutron star physics. 
Since future observational campaigns will not only provide mass, radius and rotation rate measurements, but also neutron star oscillation modes, a unified framework to analyze the joint knowledge becomes necessary. While several attempts exist in the literature, the here presented framework explicitly takes into account rotational effects by using state-of-the-art universal relations for their oscillation spectra. To demonstrate its performance we apply it to observables that take into account rotational effects to best possible precision and use our independent model for the reconstruction. Only a few high precision neutron star measurements are required to put tight constraints on the underlying equation of state.
\end{abstract}

\maketitle

\section{Introduction}

The era of multi-messenger astronomy promises many exciting new insights into unsolved problems. Among the currently most actively studied ones is the determination of the nuclear equation of state (EOS) under extreme conditions, which is expected to be realised in nature only inside neutrons stars. Current predictions coming from nuclear theory do not provide a unique answer, but a wide range of so-called realistic EOS. Combining such predictions with terrestrial nuclear physics experiments allows to put constraints on the possible allowed range, although the most extreme conditions much above nuclear saturation density cannot be accessed directly. Existing indirect constraints come from astronomical observations of neutron star properties, e.g., the famous two-solar-mass neutron star \cite{2010Natur.467.1081D} or their masses and radii by modeling hot spots using NICER observations \cite{Miller:2019cac,Riley:2019yda}. With the beginning of gravitational wave astronomy by the LIGO and Virgo observatories, binary neutron star mergers can be observed too and require a solid understanding of the strong and dynamical regime of general relativity \cite{2017PhRvL.119p1101A}. Combining gravitational waves with the many windows of traditional astronomical observations, allows to jointly explore nuclear physics and gravity in the multi-messenger approach \cite{2017ApJ...848L..12A,2017ApJ...848L..13A,2017ApJ...850L..19M,2017ApJ...850L..34B,2018ApJ...852L..29R,2018PhRvL.121p1101A,2017PhRvL.119y1303S,2017PhRvL.119y1304E,Most:2018hfd,Pacilio:2021jmq,Biswas:2021paf}. With improving performance of existing gravitational wave detectors and next generation detectors, e.g., the Einstein Telescope \cite{Maggiore:2019uih}, observations of neutron star oscillations will become available sooner or later as well. 

In this work we demonstrate how the nuclear EOS can be constrained to pristine accuracy by future measurements of bulk properties and $f$-modes of slowly rotating neutron stars.
Reconstructing or constraining the EOS from actual---or for the moment, mock---observations (the so-called \emph{inverse problem}) has first been addressed in a mathematically rather rigorous approach by Lindblom and Indik~\cite{1992ApJ...398..569L, 2012PhRvD..86h4003L, Lindblom:2014sha, 2014PhRvD..89f4003L} for the case that masses and radii of neutron stars have been measured. This approach has soon been extended by Andersson and Kokkotas to also take oscillation modes into account, thereby giving rise to the new field of \emph{gravitational wave asteroseismology} \cite{1996PhRvL..77.4134A, 1998MNRAS.299.1059A}; rapidly rotating stars were accounted for in the Cowling approximation \cite{2013PhRvD..88d4052D, 2015PhRvD..92l4004D}. Following the introduction of a piecewise polytropic approximation for the EOS (which we will discuss in detail below), the inverse problem has also been investigated employing a Bayesian approach via Markov chain Monte Carlo (MCMC) sampling \cite{2018PhRvD..97h4014A} and as a minimization problem on the corresponding parameter space \cite{2019PhRvD..99j4005M, 2019arXiv190308921M}. An approach avoiding parametrized EOS and their limitations, building purely on observations, has been presented in \cite{2019PhRvD..99h4049L}; further, we are aware of just one study specifically  investigating GW190814 while taking rapid rotation into account \cite{2021MNRAS.505.1600B}. However, all cited studies focus on non-rotating systems or (if incorporating oscillation modes) rely on major approximations like the Cowling approximation. In this work, we demonstrate explicitly that a slow rotation approximation combined with universal relations obtained from high precision calculations are reliably able to explore the nuclear EOS from a small set of neutron star observations. Our framework assumes that neutron star properties are known with some uncertainties from future observations and directly tackles the inverse problem from there. The general structure of the framework presented in this work can be categorized into the two main aspects of solving the direct and inverse problem. We summarize them in the following and provide full details in Sec.~\ref{direct_problem} and Sec.~\ref{inverse_problem}.

The \textit{direct problem} covers an adequate parametrization of the nuclear EOS for neutron stars for which we use the widely used piecewise polytropic parametrization proposed by Read \textit{et al.}~\cite{Read:2008iy}. Subsequently, one needs stellar properties that can (in principle/future) be obtained from observations, as well as a framework to solve the relevant equations of general relativity to compute them. For the non-rotating case, the equations of stellar structure are simply the Tolman-Oppenheimer-Volkoff (TOV) equations and other observables or bulk properties of such stars can be found by solving additional ordinary differential equations (like moment of inertia or tidal deformability). For rotating neutron stars we resort to a slow rotation approximation scheme in combination with universal relations in order to obtain estimates for the frequencies of the quadrupolar $f$-mode in an efficient way. The result of this part is a large set of tabulated data that contains all observables for neutron star sequences on a grid of EOS parameters, which corresponds to our ``EOS database''. 

The \textit{inverse problem} is about extracting the underlying EOS from observed data. This is done using a Bayesian approach via MCMC sampling after defining a suitable likelihood. For rotating stars, it will be computationally too expensive to solve the generalized TOV equations in every step of the sampling. Instead, we will use our EOS database to evaluate the likelihood efficiently. The MCMC sampling is done employing the MIT licensed python package \textsc{emcee} \cite{Foreman-Mackey:2012any}. The result of this part is the posterior distribution of the EOS parameter space for provided (simulated) observations, which ultimately defines our reconstructed knowledge of the nuclear EOS of neutron stars from future observations.

In Sec.~\ref{application_results} we show our results when assuming different sets of hypothetically observed neutron star observables. We find that the slow rotation framework in combination with universal relations for generally rotating neutron stars works reliable for moderate rotation rates. Finally, we discuss our results and possible extensions in Sec.~\ref{discussion} and provide our conclusions in Sec.~\ref{conclusions}. Additional results mentioned in the main text can be found in the Appendix~\ref{appendix}.

\section{Direct Problem}\label{direct_problem}

In this section we outline the different aspects of the direct problem. We start with the EOS in Sec.~\ref{direct_problem_eos}, discuss the computation of neutron star models in Sec.~\ref{direct_proble_models}, introduce the slowly rotating framework in Sec.~\ref{slowly_rotating_framework} and outline the structure of our database in Sec.~\ref{direct_proble_data}.

\subsection{EOS Representation}\label{direct_problem_eos}

Since the true nuclear EOS is currently not known and the possible range of so-called realistic EOS is still large, we have decided to work with a parametrized EOS framework. A suitable parametrization must at least be able to cover efficiently and with high accuracy all known and still allowed realistic EOS. Much work on this problem has been presented in the literature (see, e.g., Refs. \cite{Read:2008iy, 2010PhRvD..82j3011L, 2020PhRvD.102h3027O}). Since matter in neutron stars is usually assumed to have a much smaller temperature than the Fermi temperature, the chemical potential can be neglected and the pressure $p$ only depends on the rest-mass density $\rho$. 

A commonly used analytic approximation for this situation is to match the free parameters of a polytropic EOS with microscopic/nuclear theory computations. A polytropic EOS is given by
\begin{align}\label{polytropiceos}
    p(\rho) = K \rho^\Gamma,
\end{align}
where $\rho$ describes the rest-mass density, $\Gamma$ is the adiabatic index, and $K$ some scaling factor. Energy density $\epsilon$ and rest-mass density $\rho$ are then linked via the first law of thermodynamics, $\rho \dif\epsilon = (\epsilon + p) \dif\rho$. In order to provide a better representation of realistic EOSs, Eq.~\eqref{polytropiceos} can be generalized to be a piecewise polytropic EOS
\begin{align}\label{polytropiceos2}
    p(\rho) = K_i \rho^{\Gamma_i},
\end{align}
which is defined on a series of rest-mass density intervals $[\rho_{i-1}, \rho_i]$ and $K_i$ and $\Gamma_i$ are the respective parameters. 
In this work, we follow the parametrization introduced by Read \textit{et al.}~\cite{Read:2008iy}, in which a piecewise polytropic approximation (henceforth PPA) is fitted to a realistic EOS using four free parameters. The model has two fixed rest-mass density steps at $\rho_1 = 10^{14.7}\,\mathrm{g/cm^3}$ and $\rho_2=10^{15}\,\mathrm{g/cm^3}$. The free parameters are three adiabatic indices $\Gamma_1,\Gamma_2,\Gamma_3$ and the pressure $p_1$ at the density $\rho_1$. For a visualization and a list of PPA coefficients for a large number of realistic EOS, we refer to Fig.~2 and Tab.~III in Ref.~\cite{Read:2008iy}; more PPA coefficients for other EOS can be found in Ref.~\cite{2019PhRvD..99l3026K}.

By providing a framework to find the best matching EOS parameters for a given realistic EOS, Read~\textit{et al.} have demonstrated that this parametrization reproduces multiple neutron star properties of realistic EOS at percent level precision or better. We therefore expect that using this EOS as model for the inverse problem will be sufficient to recover general, realistic nuclear EOS at percent level.

\subsection{Neutron Star Equilibrium Configurations}
\label{direct_proble_models}

For a given set of EOS parameters $\theta = (p_1, \Gamma_1, \Gamma_2, \Gamma_3)$, it will become necessary to generate a sequence of neutron star models with varying central energy density (resulting, e.g., in the usual mass-radius-diagram) in order to compare them to the observed data. To do so, the equations governing the stellar structure need to be solved. Throughout this work, we use general relativity, which yields the well-known TOV equation \cite{PhysRev.55.364,PhysRev.55.374} (for the non-rotating case)
\begin{align}\label{TOVeq1}
    \frac{\text{d}p}{\text{d} r} = - \frac{Gm}{r^2} \epsilon \left(1+\frac{p}{\epsilon c^2} \right) \left(1 + \frac{4 \pi r^3 p}{mc^2}\right) \left(1-\frac{2Gm}{rc^2} \right)^{-1},
\end{align}
where the mass $m$ contained inside the ball with radius $r$ is given by
\begin{align}\label{TOVeq2}
    m(r) = 4 \pi \int_{0}^{r} {r^\prime}^2 \epsilon \text{d}r^\prime.
\end{align}
This system of equations is closed once an EOS $p(\rho)$ has been chosen. As usual, we specify a central energy density and then integrate the TOV equations from the center of the star to its surface which is defined as the location where the pressure first vanishes.

Along with the TOV equation, we solve Hartle's equation (cf. Eq. (46) in Ref. \cite{1967ApJ...150.1005H}) in order to access the moment of inertia $I$ of the neutron star. We need this quantity later in the universal relations for the $f$-mode frequency.

\subsection{Accounting for Slowly Rotating Configurations}
\label{slowly_rotating_framework}

While the TOV equation is comparably easy to solve, the situation becomes significantly more complex if we want to include rotating equilibrium configurations. Generating those neutron star models is not a technical issue (e.g., the \textsc{rns} code reliably solves the relevant equations \cite{1995ApJ...444..306S, 1998A&AS..132..431N, rns-v1.1}), however, it is computationally considerably more expensive and for simplicity, we will restrict ourselves to the bulk properties of non-rotating configurations in this study.

This does not mean that we ignore the rotation of neutron stars entirely: We artificially furnish the non-rotating configurations with an angular rotation rate $\Omega$\footnote{Angular rotation rate and spin frequency of the star are linked via $\Omega = 2\pi f_{\rm spin}$.}; as mass and radius are affected only to quadratic order in $\Omega$ \cite{1967ApJ...150.1005H}, we may do this as long as we ensure that we allow only observations of sufficiently slowly rotating neutron stars. We will discuss a criterion for the maximally allowed rotation rate so that changes in mass and radius are negligible below in the Sec.~\ref{sec:omega_limit}.

In order to link rotation rate with the neutron star bulk properties, we also want to employ the frequencies of the co- and counter-rotating quadrupolar $f$-mode of that neutron star model; we denote with $\sigma^{\rm s}$ the frequency of the stable branch (co-rotating) and with $\sigma^{\rm u}$ the frequency of the potentially unstable branch (counter-rotating). Those frequencies are affected already at linear order in $\Omega$, i.e., we cannot use the $f$-mode frequency of the non-rotating star. A direct determination of the $f$-mode frequency of rotating neutron star models in full general relativity is a highly involved task \cite{PhysRevD.102.064026}, which is why we will resort to estimating those frequencies via a universal relation that has been proposed in earlier work \cite{Kruger:2019zuz}. This universal relation has the form
\begin{align}\label{universal_relation}
    \hat{\sigma}^i =  \left(c^{i}_1 + c^{i}_2 \hat{\Omega} + c^{i}_3\hat{\Omega}^2 \right) +  \left(d^{i}_1 + d^{i}_3 \hat{\Omega}^2 \right) \eta.
\end{align}
where the different quantities are defined as follows: The index $i$ is to distinguish between the co- and counter-rotating branch of the $f$-mode; further, we have $\hat{\sigma}^{i} = \bar{M} \sigma^{i}/\mathrm{kHz}$ and $\hat{\Omega} = \bar{M} \Omega/\mathrm{kHz}$, where $\bar{M} = M/M_\odot$. The effective compactness $\eta$ is related to the mass and moment of inertia $I$ of the star via $\eta = \sqrt{\bar{M}^3/I_{45}}$, with $I_{45} = I/10^{45} \mathrm{g \, cm^2}$. The numerical values of the coefficients $c^i$ and $d^i$ have been reported in the same work~\cite{Kruger:2019zuz} and additional information can be found in Ref.~\cite{Kruger:2021zta}.

\subsection{The EOS database}
\label{direct_proble_data}

In this study, we will assume that an observation $D$ is a tuple of $N_O = 5$ observables, i.e.,
\begin{align}\label{data_vector}
    D = \left( M, R, \Omega, \sigma^{\rm s}, \sigma^{\rm u} \right).
\end{align}
In order to calculate the likelihood of accepting a proposed step in the MCMC method (cf. Sec.~\ref{inverse_problem_likelihood}), we need to know all those quantities also for our equilibrium models. As described above, solving the TOV equation along with Hartle's equation provides us with $M$, $R$ and $I$, and after assigning a series of values for $\Omega$ to a specific model, we can estimate the according $f$-mode frequencies via the universal relation.

As we want our MCMC walkers to explore the EOS parameter space, i.e., along a chain of values for $\theta = (p_1, \Gamma_1, \Gamma_2, \Gamma_3)$, we need knowledge of all equilibrium models that belong to a particular EOS given by $\theta$. One possibility to find those equilibrium models would be to calculate the mass-radius-curves on the fly for precisely the value of $\theta$ that is required by the proposed MCMC step and equip those models with the $f$-mode frequencies for some values of $\Omega$. This approach is computationally rather expensive since it is quite likely that none of the MCMC walkers will precisely cross their paths which means that none of the hitherto computed mass-radius-curves can be used for a likelihood calculation another time. This issue could be overcome by using a mass-radius-curve of an already calculated EOS $\theta'$ that is ``sufficiently'' close to the desired EOS $\theta$.

As using EOS with approximated parameters is inevitable, we opt to build a large database of mass-radius-curves at a set of specific points $\left\{ \theta_n \right\}$ in the EOS parameter space. When an MCMC walker needs to calculate the likelihood of making a step to point $\theta'$ in the parameter space, this likelihood will be calculated with respect to the EOS $\theta_{\rm approx}$,  where $\theta_{\rm approx}$ is the EOS present in the database that is closest to $\theta$ (in the sense of being the nearest neighbour).

Our approach in building the EOS database is to be rather agnostic, i.e., we choose to enlarge the parameter space and let the MCMC analysis find the ``best'' fitting EOS to the observations, rather than confining our database by various astrophysical constraints. As we employ the usual piecewise polytropic parametrization by Read~\textit{et~al.}, we refer to their Tab. III in Ref.~\cite{Read:2008iy} and simply assume intervals for $p_1$, $\Gamma_1$, $\Gamma_2$, and $\Gamma_3$ such that all provided best fits are covered; the intervals we choose for the parameters are
\begin{align}
    \log_{10}(p_1) & \in [33.940, 34.860 ], \\
    \Gamma_1  & \in [ 2.000, 4.070 ], \\
    \Gamma_2  & \in [ 1.260, 3.800 ], \\
    \Gamma_3  & \in [ 1.290, 3.660 ],
\end{align}
where $p_1$ is given in $\text{dyn}/\text{cm}^2$. We divide each of those four intervals onto a grid of 50 evenly spaced grid points; at each grid point of that four-dimensional parameter space, we calculate the mass-radius-curve as described above. This results in principle in $6.25 \times 10^6$ mass-radius-curves to be calculated.

We are aware that our chosen parameter space is sizeable and contains a large number of EOSs that could be ruled out even without applying modern astrophysical constraints. In order to reduce the size of the database somewhat, we apply the lax constraint\footnote{A two-solar-mass neutron star has been discovered in 2010 \cite{2010Natur.467.1081D} which in principle rules out all EOS that result in a maximum mass model with $M_{\rm max} \lesssim 2.0\,M_\odot$.} that the EOS must be able to support neutron stars of at least $M \ge 1.8 M_\odot$; this eliminates $\approx 2.95 \times 10^6$ candidates from our database. Furthermore, we strictly limit our mass-radius-curves at the lower mass end to $M \ge 1.17\,M_\odot$.

\subsection{Validity of the Slow Rotation Approximation}
\label{sec:omega_limit}

In order to quantify the expected precision of the non-rotating model when used to approximate the bulk properties $M,R$ of rotating neutron stars, we define the following dimensionless error
\begin{align}\label{delta_error}
\delta = \sqrt{\delta_R^2 + \delta_M^2},
\end{align}
which implicitly depends on the underlying EOS. Here the relative errors for the mass and radius are defined as
\begin{align}
\delta_R &= \frac{|R^0(M) - R^\Omega(M)|}{R^0(M)},
\\
\delta_M &= \frac{|M^0(R) - M^\Omega(R)|}{M^0(R)}.
\end{align}
The tuple $[R^0, M^0]$ simply refers to a given point on the mass radius curve in the non-rotating case, while $[R^\Omega, M^\Omega]$ describes the tuple on the rotation dependent mass radius curve that has the smallest distance to the non-rotating tuple $[R^0, M^0]$. When computed for many points on the non-rotating mass radius curve with respect to many rotating mass radius curves, one can quantify the expected precision of the model as function of rotation. This definition does not make any underlying assumption about the central density or axis ratio.\footnote{We note that by axis ratio we mean the ratio of the coordinate radii $r_p$ at the pole and $r_e$ at the equator; this is purely out of simplicity as this ratio is heavily involved in the employed \textsc{rns} code. However, at the axis ratios that we consider here ($r_p/r_e \gtrsim 0.98$), the difference between the ratio of the coordinate radii and that of the proper radii is negligible.}

\begin{figure}
\centering
\includegraphics[width=1.0\linewidth]{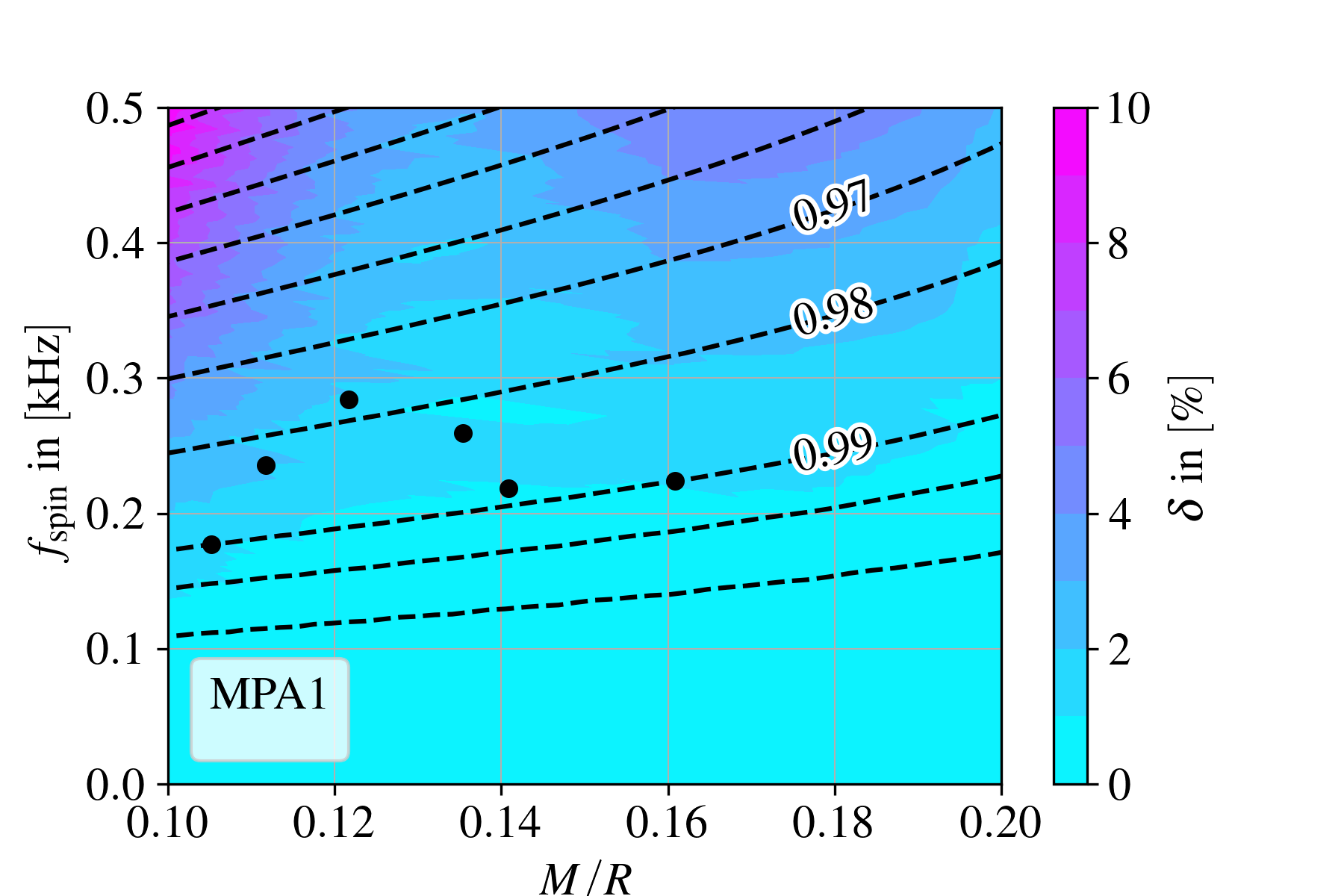}
\caption{The overall relative error $\delta$ defined in Eq.~\eqref{delta_error} for rotating stars on the compactness $M/R$ vs rotation rate $f_{\rm spin}$ parameter space for the MPA1 EOS. The relative error is shown in percent and color coded. The black dots show our data points listed in Table~\ref{table_1} and used in Sec.~\ref{application_results}. The black dashed lines show sequences of fixed axis ratio of $0.996, 0.993, 0.99$ and subsequently decreasing in steps of $0.01$ (we show three axis ratios as inline labels for clarity).}
\label{EOS_delta_1}
\end{figure}

\begin{figure}
\centering
\includegraphics[width=1.0\linewidth]{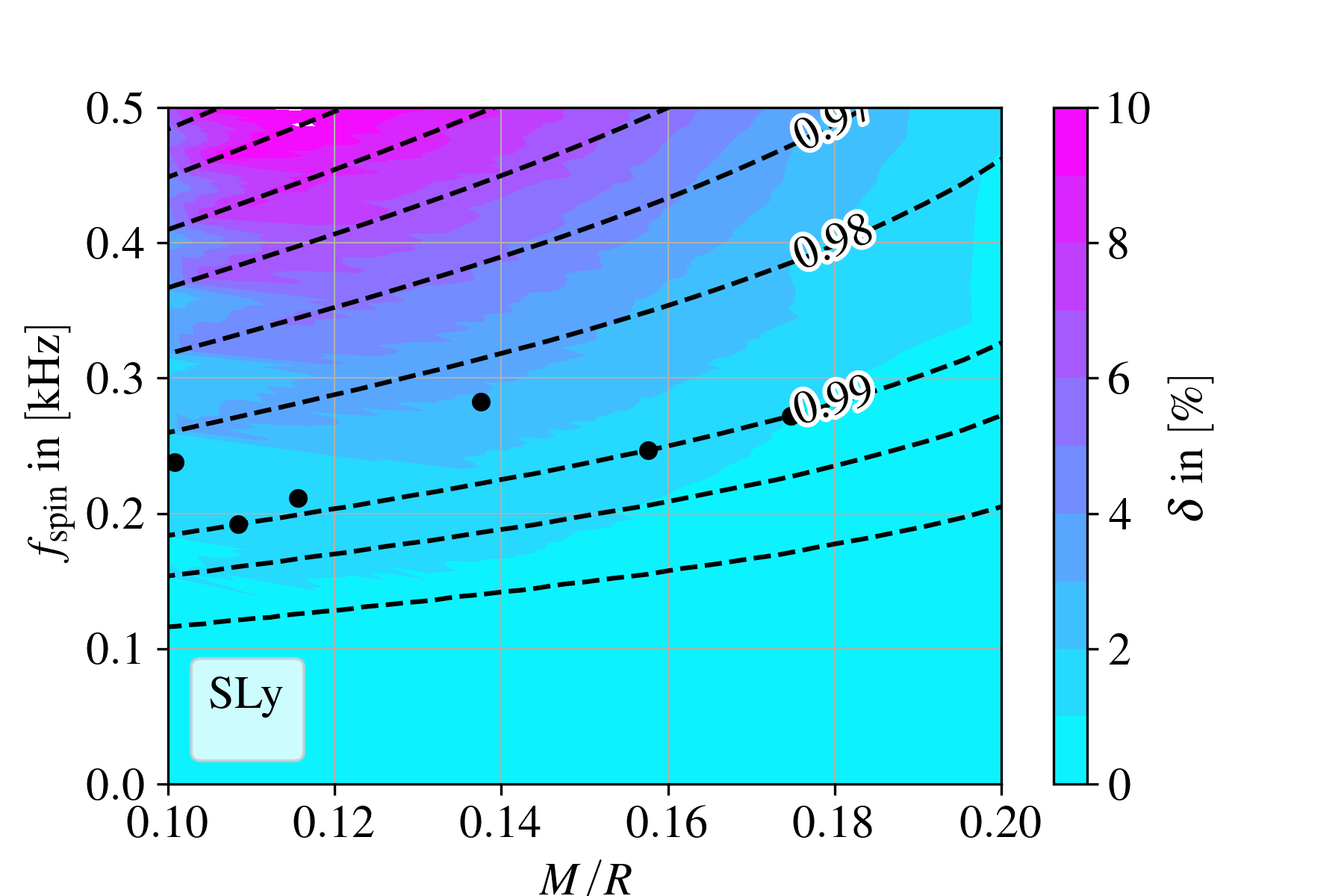}
\caption{Same as Fig.~\ref{EOS_delta_1} but for the SLy EOS and the black dots show our data points listed in Table~\ref{table_2}.}
\label{EOS_delta_2}
\end{figure}

However, since the rotation rate itself is not the most suitable measure for the slow rotation approximation, but rather the axis ratio between polar and equatorial radius, we have computed $\delta$ along multiple mass-radius-curves computed for fixed axis ratios for different EOS. The results for $\delta$ as function of compactness and rotation rate can be found in Fig.~\ref{EOS_delta_1} for the MPA1 EOS and in Fig.~\ref{EOS_delta_2} for the SLy EOS. The color code shows the value of $\delta$ in percent and lines of constant axis ratio are indicated by black dashed curves. We also added the simulated observation data points that we will use in Sec.~\ref{application_results} for the inverse problem as black dots. Their full details can be found in Table~\ref{table_1} and Table~\ref{table_2}. 

We have also computed $\delta$ for other realistic EOS and find that the slow rotation approximation works well as long as the axis ratio is larger than $0.98\sim0.99$. As expected there is an overall trend that large rotation rates make the non-rotating assumption invalid, thus increasing $\delta$, but the quantitative details vary across different EOS. It is not even qualitatively universal how $\delta$ depends on the compactness and rotation rate for a given EOS as soon as one considers axis ratios smaller than $0.96\sim0.97$.

Since one does not know the true underlying EOS in the inverse problem, it is not always obvious which observations can be interpreted with the slow rotation approximation and which cannot. As the results in this section show, there are indeed suitable (EOS dependent) regions of the parameter space that allow for values of $\Omega$ that are not small, but still fall within the approximation. Depending on the desired precision, we expect that the slow rotation approximation holds on percent level for stars with rotation rates below $\sim 200\,\mathrm{Hz}$ and around $10\,\%$ for rotation rates below $\sim 400\,\mathrm{Hz}$. For very compact configurations the overall error seems to be systematically smaller. This is expected since more compact configurations should require a larger rotation rate to be deformed by the same amount than a less compact star.

\section{Inverse Problem}\label{inverse_problem}

In this section we outline our framework to solve the inverse problem. We give a brief overview of Bayesian analysis in Sec.~\ref{inverse_problem_bayesian}, introduce our likelihood in Sec.~\ref{inverse_problem_likelihood} and discuss its practical evaluation in Sec.~\ref{sec:workflow_likelihood}.

\subsection{Bayesian Analysis}\label{inverse_problem_bayesian}

Bayesian analysis allows to quantify our knowledge of the parameters of a model $\theta$ for given observations or data $D$. Its central piece is Bayes' theorem which connects the two via
\begin{align}
\mathcal{P}\left(\theta | D \right) = \frac{ \mathcal{P} \left(D | \theta \right) \mathcal{P}\left(\theta \right)}{\mathcal{P}\left( D \right)}.
\end{align}
Here $\mathcal{P}\left(\theta | D \right)$ is the posterior distribution and describes the probability of the parameters given the data. The theorem states that it is equal to the likelihood $\mathcal{P} \left(D | \theta \right)$, which describes the probability of the data given the parameters, times the prior $\mathcal{P}\left(\theta \right)$, which captures the probability of the parameters before looking at the data. The evidence $\mathcal{P}\left( D \right)$ is the probability of the data itself and acts effectively as normalization.\footnote{For a tutorial on Bayesian analysis, see, e.g., \cite{sivia2006data}.} In this work the priors of the EOS parameters $\theta = (p_1, \Gamma_1, \Gamma_2, \Gamma_3)$ are assumed to be flat and defined on intervals listed in Sec.~\ref{direct_proble_data}. The precise form of the data $D$ will be defined in Sec.~\ref{inverse_problem_likelihood}.

The actual computation of the posteriors is done using a MCMC analysis based on the MIT licensed python package \textsc{emcee} introduced in Ref.~\cite{Foreman-Mackey:2012any}. The great advantage of using MCMC methods is that one samples the posterior distribution and only requires the explicit knowledge of the likelihood and prior, but not the evidence. The latter one is usually not known a priori. The practical limitation of MCMC methods is that they can easily become computationally expensive, in particular when the parameter space is large and the posterior multi-modal. In our work the number of parameters is moderate and the computational cost using the EOS database affordable.

\subsection{Definition of Likelihood}
\label{inverse_problem_likelihood}

In order to use MCMC methods, we have to compute the likelihood $\mathcal{P}\left(D|\theta \right)$ given a set of $N_D$ observations $D = \left\{ D^i \right\}_{i=1}^{N_D}$ and an EOS given by the parameters $\theta$, where we assume that each observation $D^i$ denotes a tuple of $N_O$ individual observables $D^i = \left( D^i_j \right)_{j=1}^{N_O}$ as explained in Eq.~\eqref{data_vector}. Since our study focuses mainly on future experiments and proof of principle solutions to the inverse problem, we assume a simple likelihood of the following form
\begin{align}\label{likelihood}
\log\left( \mathcal{P} (D| \theta)\right)  
= -\frac{1}{2} \sum_{i=1}^{N_D} \sum_{j=1}^{N_O}\left( \frac{D_{j}^{i} - M_{j}^{i}(\theta)}{\bar{\sigma}^{i}_j} \right)^2,
\end{align}
as a measure as to how well our observations $D$ are described by the EOS $\theta$. 
Here, $D_j^i$ is the $j$-th observable of the $i$-th neutron star observation data point, $\bar{\sigma}_j^i$ the associated uncertainty and $M_{j}^{i}(\theta)$ the corresponding model value, the determination of which we will explain in more detail in the following Sec.~\ref{sec:workflow_likelihood}. For simplicity, and in the absence of an educated guess, we assume that the data is uncorrelated and Gaussian distributed. While measurements of neutron star observables $D^{i}$ of different neutron stars are truly independent, there usually are correlations between the individual observables, e.g., mass and radius of the same neutron star are in general not measured independently of each other. Once such correlations are known from future observations, they could as a first step be taken into account by considering suitable correlated Gaussian noise. However, we emphasize that our work demonstrates how rotational effects can be explicitly treated in the inverse problem, rather than forecasting the details of specific astrophysical observations, which is an important but non-trivial problem in itself.

\subsection{Workflow for Computing Likelihood}
\label{sec:workflow_likelihood}

The likelihood defined in Eq.~\eqref{likelihood} is not explicit about how the model values are being computed. In fact, it is non-trivial to find the model values for a given choice of the EOS parameters $\theta$, because the problem we face is parametric. In most optimization problems one provides the model values in the form $y_\mathrm{model} = f(\theta, x)$. To clarify, let us consider the simplest case in which we have measurements for $M$ and $R$ and want to determine the best EOS parameters $\theta$. Every given choice of $\theta$ yields a mass radius curve. How does this yield the model values of $M$ and $R$? We define the proposed model values as the point on the mass radius curve with the shortest geometrical distance to the data points. When we have multiple pairs for $M$ and $R$ as observation, we can repeat the procedure for each of them individually. 

The procedure qualitatively summarized here adds a minimization problem every time the likelihood is computed. In practice we determine the model values by evaluating the mass radius curve on a grid with sufficient resolution at every point, compute the geometrical distance, and choose the one with the smallest as model value. The evaluation of the curve is simply done by reading the corresponding template file for this particular choice for the EOS parameters. As described in Sec.~\ref{direct_proble_data}, the EOS database has been computed once and is therefore readily available during the MCMC sampling. The concept of providing model values can easily be generalized when the dimension of each data vector is not $2$, but larger and includes more observation types. 

In the more general case one has to find all the proposed model values for the given EOS parameters. To do so we compute the shortest distance of each observed neutron star data vector $D^{i, \mathrm{data}}$ to the multidimensional curve of the proposed EOS parameters defined by (for better readability we suppress the index $i$ in the next two equations)
\begin{align}\label{distance1}
\Delta^2_{k_\mathrm{min}} \equiv \mathrm{min} \left(\Delta^2_k \right)
\end{align}
where $k\in N_\mathrm{seqgrid}$ labels the values for the discretized EOS sequence and $\Delta_k^2$ is defined as
\begin{align}\label{distance2}
\Delta_k^2 \equiv \sum_{j =0}^{N_\mathrm{dim}}
\left(D^\mathrm{k, model}_j - D^\mathrm{data}_j \right)^2.
\end{align}
The multidimensional point on the EOS curve labeled by $k_\mathrm{min}$ then corresponds to the shortest distance, which defines the model proposal $M_{j}^{i}$. This procedure is repeated for all neutron star data points. With the set of all proposed model values at hand, one can finally compute the likelihood Eq.~\eqref{likelihood}.

When considering rotation, we compute the $f$-mode frequencies on a grid with the same resolution as the other observables from the universal relation. Note that this is not part of the EOS database, because the computation via universal relation is very fast and the EOS database size would increase by the number of grid points for $\Omega$, which we set to be the same as the number of points used for each EOS curve.

\section{Application and Results}\label{application_results}

In the following we apply our framework to two different sets of simulated neutron star data obtained from either assuming the MPA1 EOS in Sec.~\ref{app_MPA1} or the SLy EOS in Sec.~\ref{app_SLy}. In both cases we obtained the bulk properties taking fully into account rotation by using the \textsc{rns} code and compute the $f$-mode frequencies using the time evolution code presented in \cite{PhysRevD.102.064026}. The data provided in this way is as ``realistic'' as currently possible, since it is based on state-of-the-art calculations. The choice of the two EOS only serves as proof of principle and can be repeated with any other viable EOS candidate. In both applications we assume that all simulated measurements are known with pristine precision of $3\,\%$ relative error and we consider the noiseless limit. This is clearly not possible with current data, but the main purpose is to demonstrate the capabilities of our framework and the underlying assumptions regarding the treatment of rotational effects. We are using parallelized MCMC sampling by initiating it on 20 CPUs, each with 100 walkers with 500 steps, and combined all chains in post processing. As burn-in we have removed the first half of each chain. The sampling takes around one hour on a workstation.

\subsection{Injection of MPA1 EOS}\label{app_MPA1}

As first example we choose the MPA1 EOS \cite{Muther:1987xaa} as underlying EOS for which the PPA parameters are given by
\begin{align}
\theta^\mathrm{MPA1} = \left(34.495, 3.446, 3.572, 2.887\right).
\end{align}
The data we created as hypothetical future observation is shown in Table~\ref{table_1}. The selected neutron stars cover a mass range between $1.311\,M_\odot - 2.002\,M_\odot$ and have rotation rates between $177\,\mathrm{Hz} - 284\,\mathrm{Hz}$. The $f$-mode splitting is clearly present for all neutron stars and the axis ratio lies between $0.978 - 0.990$. 

\begin{figure}
\centering
\includegraphics[width=1.0\linewidth]{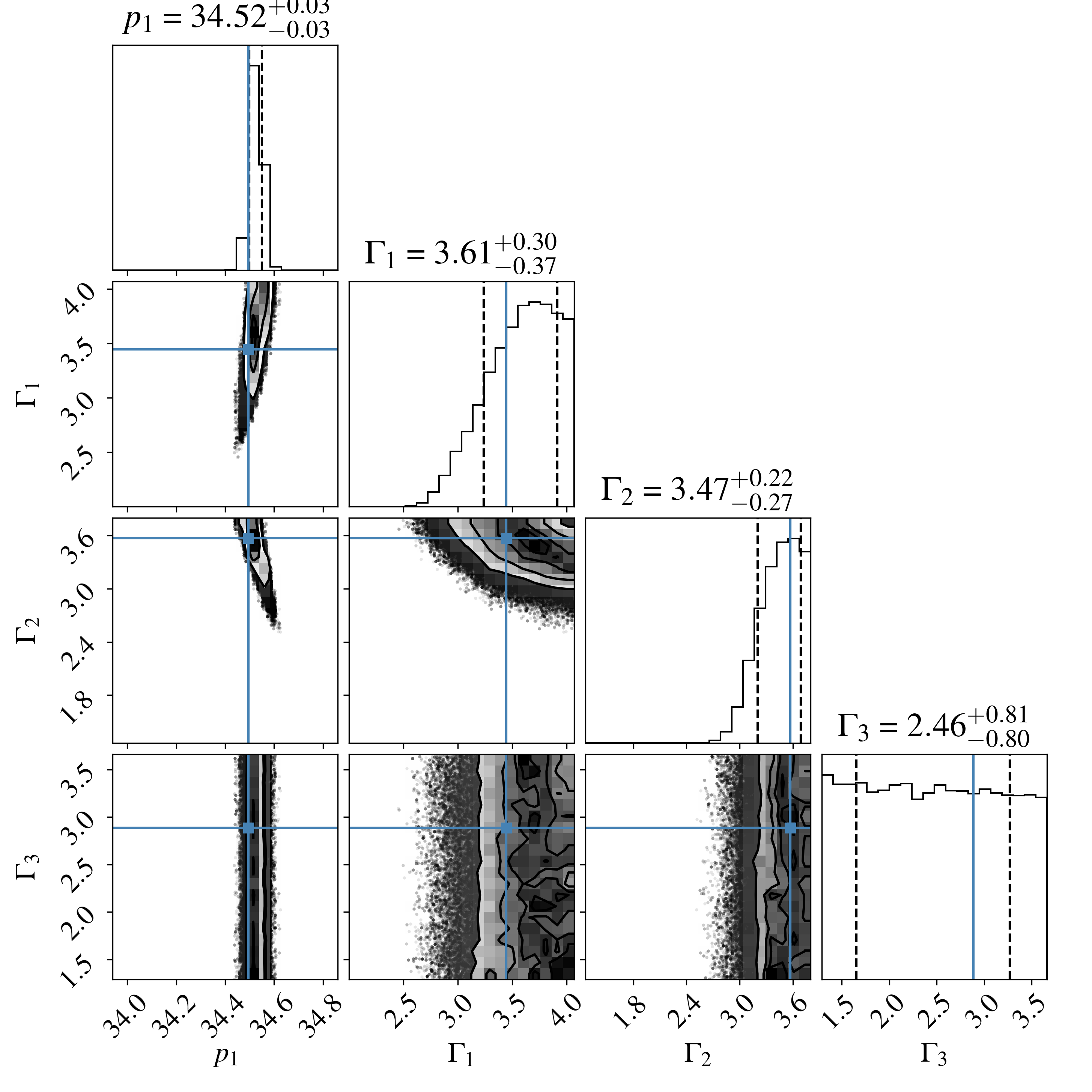}
\caption{Results of the MCMC sampling for our hypothetical data set based on the EOS MPA1 (cf.~Table~\ref{table_1}). The diagonal panels show the posterior distributions of the EOS parameters $\theta$, while the off-diagonal panels show the corresponding correlations. The quoted error ranges are for quantils of $[0.16, 0.84]$ and are shown with vertical dashed lines. The injected EOS parameters are indicated with blue crosses.}
\label{model_res1}
\end{figure}

The results of the MCMC sampling is shown in the corner plot\footnote{It was created using the Python package \textsc{corner} \cite{corner}.} in Fig.~\ref{model_res1}. The diagonal panels show the posterior distribution of each EOS parameter, while the correlations are captured in the scatter plots shown in the off-diagonal panels. It can clearly be seen that it is possible to recover the injected parameters with varying accuracy, but either within $[0.16, 0.84]$ quantils (shown with vertical dashed lines) or very close to it. While the part of the EOS described by $p_1, \Gamma_1, \Gamma_2$ can be recovered well (i.e., the EOS up to the largest dividing density $\rho_2 = 10^{15}\,\mathrm{g/cm^3}$ employed in the PPA parametrization), the high density parameter $\Gamma_3$ is basically unconstrained. A closer look into Table~\ref{table_1} shows that the selected neutron stars for the MPA1 EOS are not very heavy; indeed, the central rest-mass densities of the neutrons stars in our simulated data set ranges between $0.65 - 0.79 \times 10^{15}\,\mathrm{g/cm^3}$, i.e., below $\rho_2$.

\begin{figure}
\centering
\includegraphics[width=1.0\linewidth]{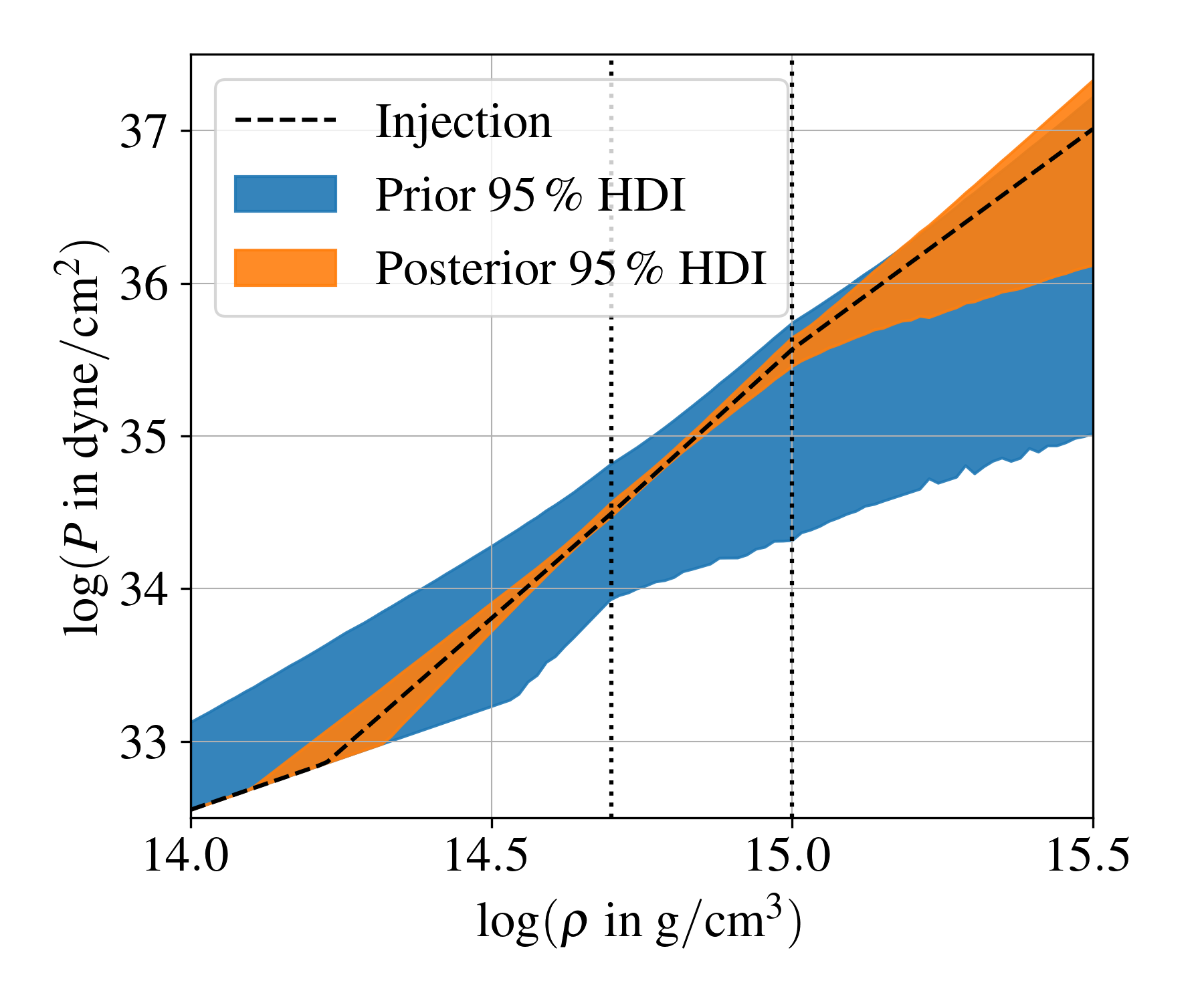}
\caption{Results for the EOS using the data shown in Tab.~\ref{table_1}. The black dashed line marks the injected EOS. We show the highest density intervals of $95\,\%$ for the samples of the prior parameter distribution (blue) and posterior distribution (orange).}
\label{model_res2}
\end{figure}

The implications of the obtained EOS parameters on the EOS is shown in Fig.~\ref{model_res2}. We compare our prior knowledge of the EOS, defined by the full parameter range contained in the EOS database, with our knowledge after performing the MCMC. To do so we sample from the prior and posterior distributions to compute the Read \textit{et al.} EOS and then compute the highest density intervals (HDI)\footnote{We used the Python package \textsc{ArviZ} \cite{arviz_2019}.} of $95\,\%$. It can be clearly seen that the injected EOS is included in the posterior distribution, which is very informative compared to the prior. The injected EOS is shown as black dashed line and falls well within the $95\,\%$ HDI of the posterior. 

\begin{table}
\[
\begin{array}{c|ccccc|c}
\toprule
n & M & R & f_{\rm spin} & \sigma^{\rm u} & \sigma^{\rm s} & \mathrm{ar}\\
 & [M_\odot] & [\mathrm{km}] & [\mathrm{kHz}] & [\mathrm{kHz}] & [\mathrm{kHz}] &  \\
\midrule
1 & 1.764 & 12.523 & 0.218 & 1.533 & 2.116 & 0.989 \\ 
2 & 1.311 & 12.466 & 0.177 & 1.468 & 1.900 & 0.990 \\ 
3 & 2.002 & 12.445 & 0.224 & 1.618 & 2.239 & 0.990 \\ 
4 & 1.400 & 12.528 & 0.236 & 1.394 & 1.993 & 0.983 \\ 
5 & 1.700 & 12.559 & 0.260 & 1.442 & 2.129 & 0.983 \\ 
6 & 1.531 & 12.584 & 0.284 & 1.361 & 2.094 & 0.978 \\ 
\bottomrule
\end{array}
\]
\caption{Simulated data used for the inverse problem in Sec.~\ref{app_MPA1}. The assumed underlying EOS is MPA1 with PPA coefficients $\theta^\mathrm{MPA1}  = \left(34.495, 3.446, 3.572, 2.887\right)$. The relative errors are discussed in the main text. The last column shows the axis ratio (ar), which we report for completeness, but which did not enter our analysis.}
\label{table_1}
\end{table}

\subsection{Injection of SLy EOS}\label{app_SLy}

As second example we choose the SLy EOS \cite{Douchin:2001sv} as underlying EOS for which the PPA parameters are
\begin{align}
    \theta^\mathrm{SLy} = \left(34.384, 3.005, 2.988, 2.851\right).
\end{align}
The corresponding hypothetical observations are shown in Table~\ref{table_2}, which shows that the here selected neutron stars cover a mass range between $1.200\,M_\odot - 1.924\,M_\odot$ and have rotation rates between $192\,\mathrm{Hz} - 283\,\mathrm{Hz}$. Also in this case the $f$-mode splitting is clearly present for all neutron stars and the axis ratio is between $0.983 - 0.990$. Compared to the former data set for the EOS MPA1, this data set includes observations of three neutron stars whose central density is larger than $\rho_2 = 10^{15}\,\mathrm{g/cm^3}$; hence, we expect to find some kind of constraint for the high density parameter $\Gamma_3$.

\begin{figure}
\centering
\includegraphics[width=1.0\linewidth]{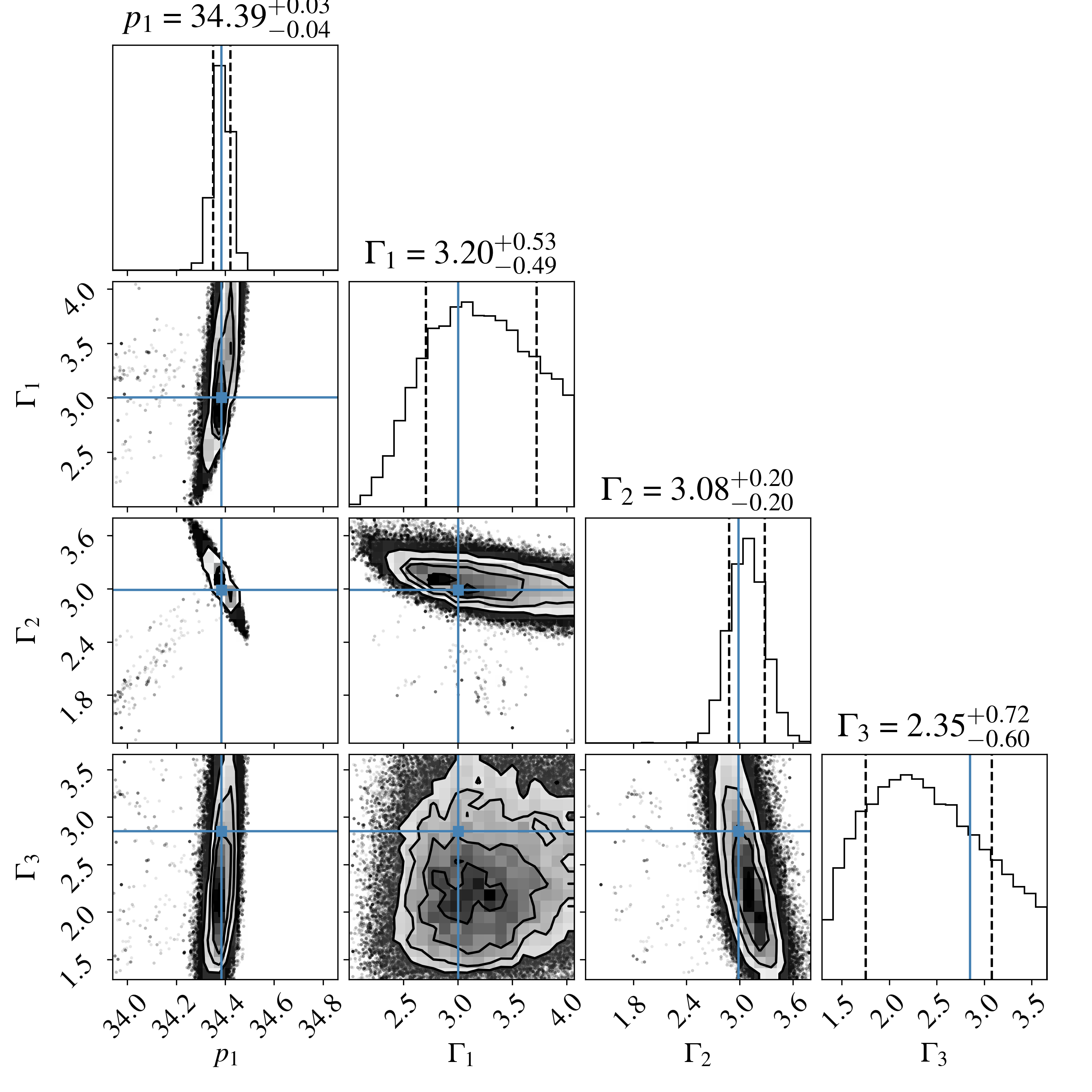}
\caption{Results of the MCMC sampling for the given data. The diagonal panels show the posterior distributions of $\theta$, while the off-diagonal panels show the corresponding correlations. The quoted error ranges are for quantils of $[0.16, 0.84]$. The injected EOS parameters are indicated with blue crosses.}
\label{file_res1}
\end{figure}

\begin{figure}
\centering
\includegraphics[width=1.0\linewidth]{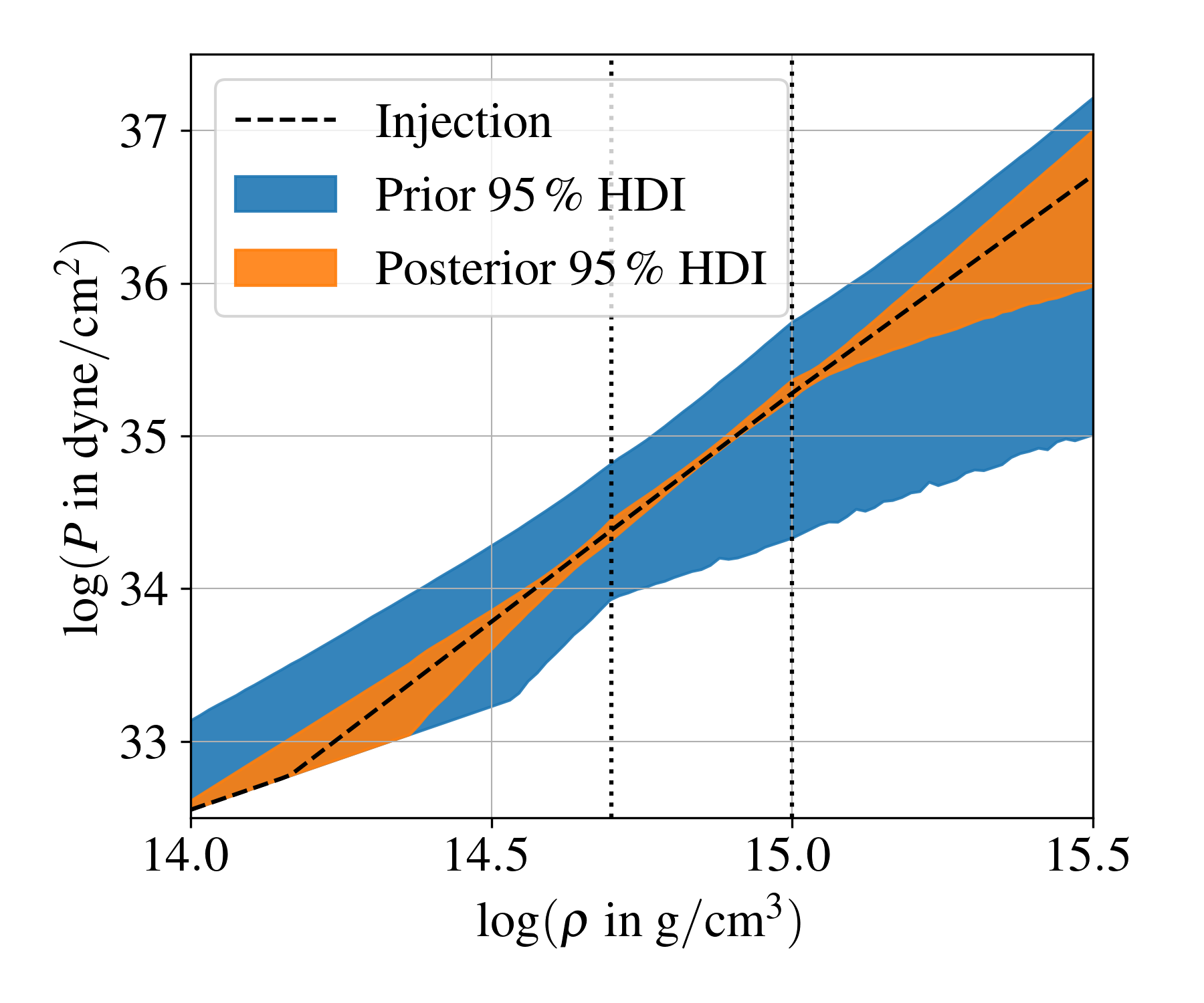}
\caption{Results for the EOS using the data shown in Tab.~\ref{table_1}. The black dashed line marks the injected EOS. We show the highest density intervals of $95\,\%$ for the samples of the prior parameter distribution (blue) and posterior distribution (orange). }
\label{file_res2}
\end{figure}

We show the results of the MCMC sampling of this case in Fig.~\ref{file_res1}, where the structure of the corner plot is the same as in Fig.~\ref{model_res1}.  In contrast to the previous case, one can now constrain all EOS parameters reasonably well. Neutron stars of the SLy EOS achieve an overall higher compactness and therefore higher central densities. This implies the influence of the high density EOS parameter $\Gamma_3$ is more relevant and thus easier to constrain. We verified that the most compact selected neutron stars have a central density slightly above the second density step $\rho_2$. In Fig.~\ref{file_res2} we show the reconstruction of the EOS along with the exact injection. It is evident that the high density region can be constrained much better compared to the MPA1 EOS.

\begin{table}
\[
\begin{array}{c|ccccc|c}
\toprule
n & M & R & f_{\rm spin} & \sigma^{\rm u} & \sigma^{\rm s} & \mathrm{ar} \\
 & [M_\odot] & [\mathrm{km}] & [\mathrm{kHz}] & [\mathrm{kHz}] & [\mathrm{kHz}] &  \\
\midrule
1 & 1.282 & 11.827 & 0.192 & 1.590 & 2.075 & 0.990 \\
2 & 1.200 & 11.902 & 0.238 & 1.486 & 2.082 & 0.983 \\
3 & 1.363 & 11.792 & 0.211 & 1.603 & 2.143 & 0.989 \\
4 & 1.924 & 11.013 & 0.272 & 1.862 & 2.649 & 0.990 \\
5 & 1.601 & 11.637 & 0.283 & 1.618 & 2.368 & 0.984 \\
6 & 1.788 & 11.340 & 0.247 & 1.798 & 2.476 & 0.990 \\
\bottomrule
\end{array}
\]
\caption{Simulated data used for the inverse problem in Sec.~\ref{app_SLy}. The assumed underlying EOS is SLy with PPA coefficients $\theta^\mathrm{SLy} = \left(34.384, 3.005, 2.988, 2.851\right)$. The relative errors are discussed in the main text.The last column shows the axis ratio (ar), which we report for completeness, but which did not enter our analysis.}
\label{table_2}
\end{table}

\section{Discussion}\label{discussion}

In the following we discuss the accuracy of the reconstructed EOS in Sec.~\ref{disc0}, the role of measurement precision in Sec.~\ref{disc1}, the validity of the slow rotation approximations in Sec.~\ref{disc2} and provide a future outlook in Sec.~\ref{disc3}.

\subsection{Accuracy of the Reconstructed EOS}\label{disc0}

When inferring the parameters of an imperfect model from realistic data one has to carefully assess the range of validity and identify possible bias in the reconstruction. Since we use ``realistic'' neutron star observables that include rotational effects, even when small for the bulk parameters, there is a finite error in the modeling. In the limit that the data would be known with pristine precision, this potentially translates into the inference of ``misleading'' posteriors. Misleading here means that the highest probability density intervals of such posteriors, e.g. the interval containing $95\,\%$ of the posterior, would not include the true parameter. Since the true parameter in a real application is not known, one would draw the wrong conclusions. 

To approximate how strong the bias based on using the slow rotation approximation and universal relations is, we have also studied cases in which the injected data has been produced from the model. One such case can be found in appendix Sec.~\ref{appendix_mock}. Here we have chosen a specific set of EOS parameters and took a sample of neutron stars from the mass radius curve. We have then chosen rotation rates similar to the ones used for the MPA1 and SLy examples and used the universal relation to compute the corresponding $f$-mode frequencies. When applying the inverse framework to this data, we find very similar posterior shapes, but in this case the injected EOS parameters are much closer to the maxima than to the $[0.16, 0.84]$ quantils, which happened for some of the parameters in the MPA1 and SLy cases. The bias described here is present when neutron star observables are known with relative errors of $3\,\%$, but becomes less relevant when the observables are known with larger uncertainties, e.g. with $5\,\%$ relative errors. We show such cases in the appendix in Sec.~\ref{appendix_5}.

\subsection{Role of Measurement Precision}\label{disc1}

We have made the simplistic assumption that future measurements may provide neutron star properties on percent level, which is clearly not justified for current or near future data. However, the main purpose of this paper is to demonstrate explicitly how the inverse problem can be solved when precise data becomes available. In contrast to the non-rotating case, where Lindblom has demonstrated how the EOS can be obtained from the mass radius curve in a non-parametric way \cite{1992ApJ...398..569L, 2012PhRvD..86h4003L, Lindblom:2014sha, 2014PhRvD..89f4003L}, a direct inversion is significantly more complicated in the rotating case. To our best knowledge, we are not familiar of any framework that has demonstrated it. Moreover, a framework that is only based on mass radius (rotation rate) does not unfold the full potential of complementary neutron star information, e.g. oscillation modes or tidal deformabilities. The Bayesian character of our framework allows easily to set any desired measurement precision for each neutron star and its properties. The assumption of Gaussian uncorrelated data in our likelihood can also be generalized to Gaussian correlated data once the correlations are known from future experiments (or simulations predicting them for experiments). Note that the main building blocks, the EOS database and the algorithm to propose the model values for a given EOS parameter, do not depend on the assumptions of the measurement precision. We provide two additional applications to cases when the relative errors are assumed to be of $5\,\%$ and $10\,\%$ in Sec.~\ref{appendix_mock}. As expected one finds less informative reconstructions, but still a great improvement compared to the priors.

\subsection{Validity of Slow Rotation}\label{disc2}

As outlined in Sec.~\ref{slowly_rotating_framework} for the direct problem, corrections to the mass and radius enter at quadratic order and we have quantified how mass, radius and rotation rate impact the expected precision. Our results for the inverse problem in Sec.~\ref{application_results} suggest that the slow rotation approximation can also be used for the inverse problem when considering neutron stars with rotation rates below $\sim 300\,\mathrm{Hz}$. In the same section we demonstrated that for some regions of the parameters a larger rotation rate can still be captured well, but this region depends on the EOS. Thus, in the inverse problem where one does not even know the axis ratio (which is a better measure for the slow rotation approximation), one is forced to set the upper bound for suitable neutron star observations rather low to avoid possible bias from using a too imprecise model. Since the universal relation for the $f$-modes is valid for arbitrary rotation, this part of the observation may be modeled correctly even for larger rotation rates and could suppress the bias even for a bit faster rotation rates. Note that the neutron star PSR J0030+0451 used in the recent NICER observations \cite{Miller:2019cac,Riley:2019yda} has a rotation rate of around $200\,\mathrm{Hz}$, which would justify the slow rotation approximation even if future measurements would be more precise.

The universal relation for the co- and counter-rotating $l=2$ quadrupolar $f$-mode has been established on percent level in Refs.~\cite{Kruger:2019zuz}. In another work \cite{Volkel:2021gke} we have used the same universal relation, but applied it to the inverse problem of single neutron stars with arbitrary rotation. By comparing EOS dependent and independent approaches, it has been verified that the universal relation can be used for the inverse problem on percent level.

\subsection{Future Extensions}\label{disc3}

Our applications demonstrate that the slow rotation framework is well suited for the inverse problem as long as neutron stars are not rapidly rotating and all EOS details are described by the Read \textit{et al.}~\cite{Read:2008iy} parametrization. However, a general analysis covering arbitrary rotation and more general EOS requires several extensions, that we want to briefly comment on in the following. 

In the presence of rotation, the mass radius curve becomes a function of the rotation and thus the EOS database has to be extended in more dimension. The calculation of equilibrium sequences of rotating neutron stars is more involved, but codes to solve the generalized TOV equations exist and are publicly available, e.g. the \textsc{RNS} code. The main problem one faces here is the increasing computational time for solving the equations, which increases the time needed to compute the EOS database by a few orders of magnitude.

The validity of the universal relations to compute the $f$-modes is not based on the slow rotation approximation, so it can still be used, but applied to the rotating neutron star equilibrium models. Future work could also include new universal relations for the damping times or other harmonics, e.g. the $l=|m|=3$ mode. More complicated is the role of tidal love numbers. To our knowledge existing works within general relativity have only been carried out in the slow rotation approximation and do not agree with each other \cite{Pani:2015hfa,Pani:2015nua,Landry:2015zfa,Landry:2017piv,Abdelsalhin:2018reg,Pani:2018inf}. 

Another aspect that one can approach with our framework is to consider modified gravity theories, in which the bulk properties are modified by additional fields, e.g. scalar fields, and study how degenerate this may impact the reconstruction of the underlying EOS. This is a non-trivial extension since viable modifications via so-called scalarization are partially constrained for non-rotating stars and the calculation of $f$-modes or other dynamical properties will be more challenging.

Finally, the Read \textit{et al.}~\cite{Read:2008iy} parametrization is just one possibility amongst many. The PPA parametrization can easily be extended (or at least structurally adopted) to explore more details of the underlying EOS or even account for more exotic possibilities, e.g., phase transitions at very high densities during binary neutron star mergers \cite{2010PhRvL.105p1102H, 2013ApJ...773...11H, Weih:2019xvw}. The framework outlined in this paper can conceptually easily be extended to EOS with more parameters, but the computational time to create the EOS database and perform the actual MCMC sampling may increase.

\section{Conclusions}\label{conclusions}

In this work we have presented a Bayesian framework that allows to reconstruct parametrized nuclear EOS from simulated observations of neutron star bulk properties and dynamical properties of slowly rotating neutron stars. We have used the four parameter parametrization of Read \textit{et al.}~\cite{Read:2008iy} to approximate a large variety of technically constructible EOS. Since bulk properties are not sensitive to small rotation rates, the dependency on the latter has been captured by including the line splitting of the dominant $l=|m|=2$ quadrupolar $f$-mode via universal relations derived for rapidly rotating neutron stars to high precision in Ref.~\cite{Kruger:2019zuz}.

To demonstrate and verify the performance of our framework we have applied it to recover two injected realistic EOS candidates given by the MPA1 and SLy models. In each case we first produced neutron star observations that include the rotational effects on the neutron star bulk properties by using the \textsc{RNS} code \cite{1995ApJ...444..306S,Nozawa:1998ak} and have computed the $f$-modes with high precision using the code of Ref.~\cite{Kruger:2019zuz}. We consider rotation rates that are of the order of pulsars and have significant effect on the $f$-modes. To justify the slow rotation approximation with our analysis in Sec.~\ref{sec:omega_limit} we have limited ourselves to rotation rates below $\sim 300\,\mathrm{Hz}$. The results presented in Sec.~\ref{application_results} clearly demonstrate that our framework with its underlying assumptions and approximations is capable of recovering the corresponding Read \textit{et al.}~\cite{Read:2008iy} parameters. 

Although this EOS parametrization is well established and widely used to represent realistic EOS instead of using tabulated data, it is not the most general model. However, it is straightforward to extend our framework to include more refined EOS parametrizations to capture more details, e.g. phase transitions at higher densities. The framework we have developed can easily take into account different precision for different neutron star observations and be extended to include additional modes or damping times when available. The absence of specific measurements, e.g. one $f$-mode is not measured, can easily be taken into account by assuming large errors. We are currently extending the framework beyond the slow rotation assumption for the bulk properties. If future gravitational wave observations provide precise measurements of neutron stars, the underlying EOS can be recovered with high accuracy.

\begin{acknowledgments}
The authors want to thank Kostas D. Kokkotas, Enrico Barausse, Marco Crisostomi, Kostas Glampedakis and Andreas Bauswein for useful discussions. SV wants to thank the University of T\"ubingen for hospitality. SV acknowledges financial support provided under the European Union's H2020 ERC Consolidator Grant ``GRavity from Astrophysical to Microscopic Scales'' grant agreement no. GRAMS-815673. This work was supported by the EU Horizon 2020 Research and Innovation Programme under the Marie Sklodowska-Curie Grant Agreement No. 101007855. CK wishes to thank SISSA and IFPU for hospitality. CK acknowledges financial support by DFG research Grant No. 413873357.
\end{acknowledgments}

\appendix

\section{Additional Results}\label{appendix}

\subsection{Application to Model Data}\label{appendix_mock}

\begin{widetext}

\begin{figure}

\begin{minipage}{1.0\linewidth}

\begin{minipage}{0.47\linewidth}
\includegraphics[width=1.0\linewidth]{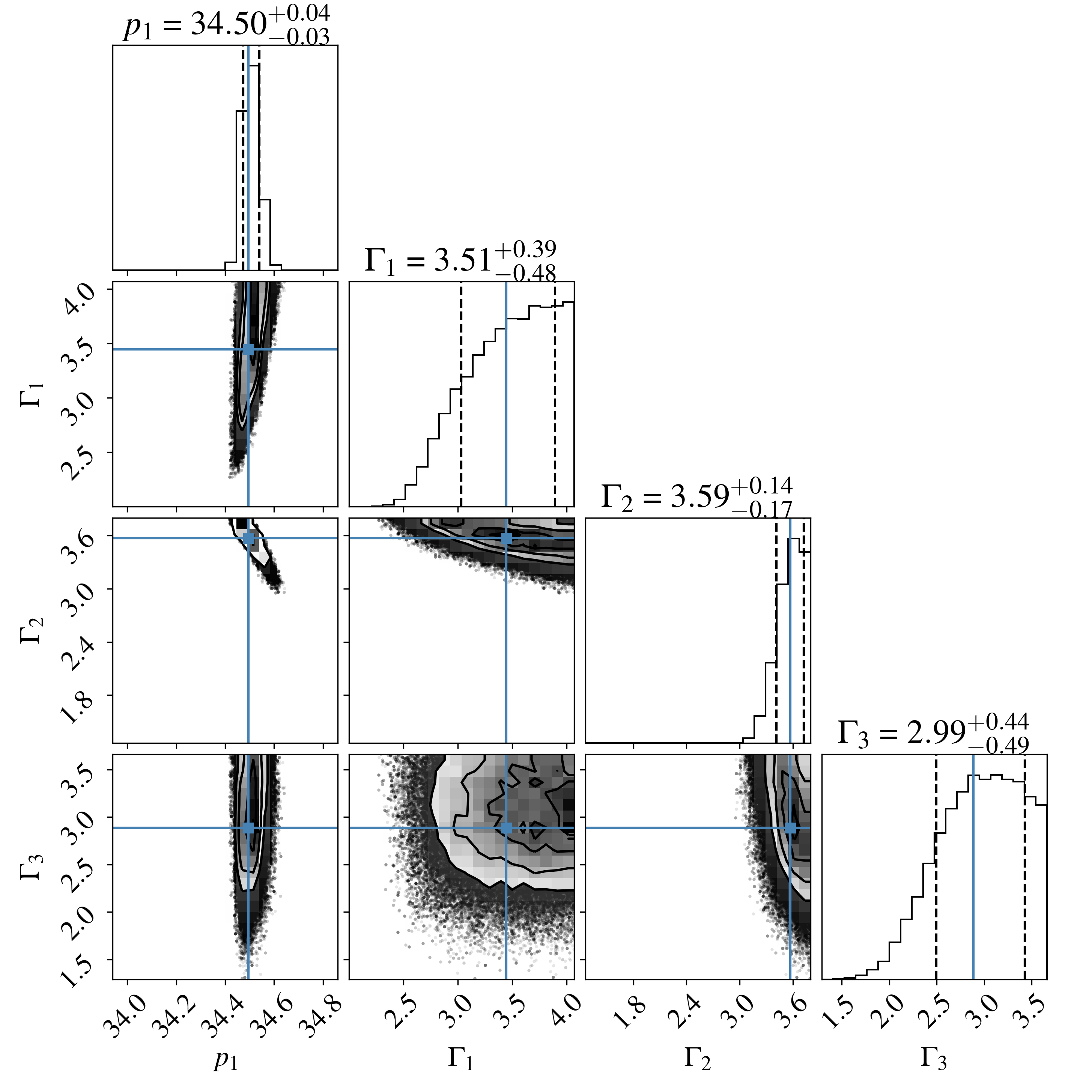}
\\~\\
\includegraphics[width=1.0\linewidth]{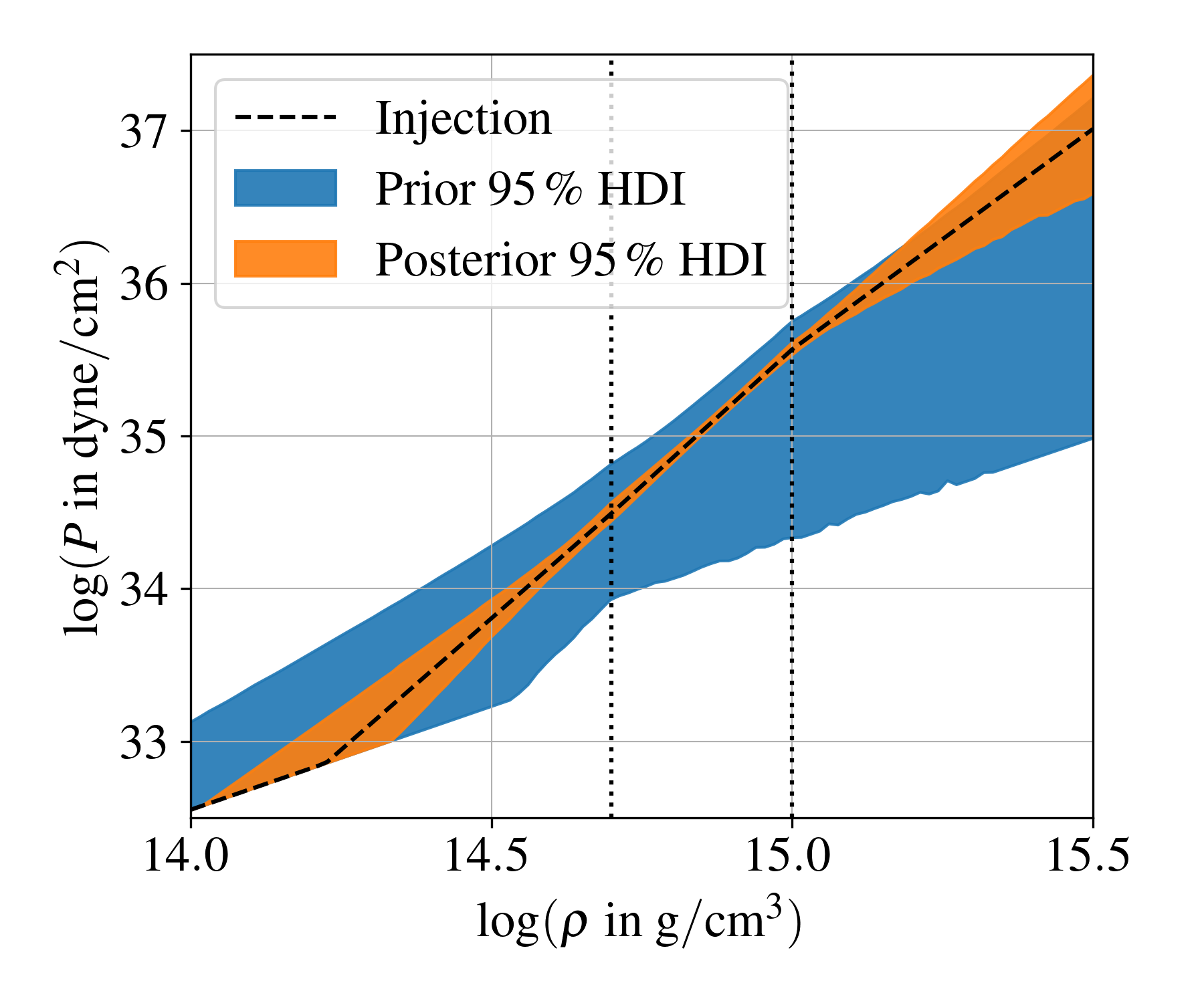}
\end{minipage}
\hfill
\begin{minipage}{0.47\linewidth}
\includegraphics[width=1.0\linewidth]{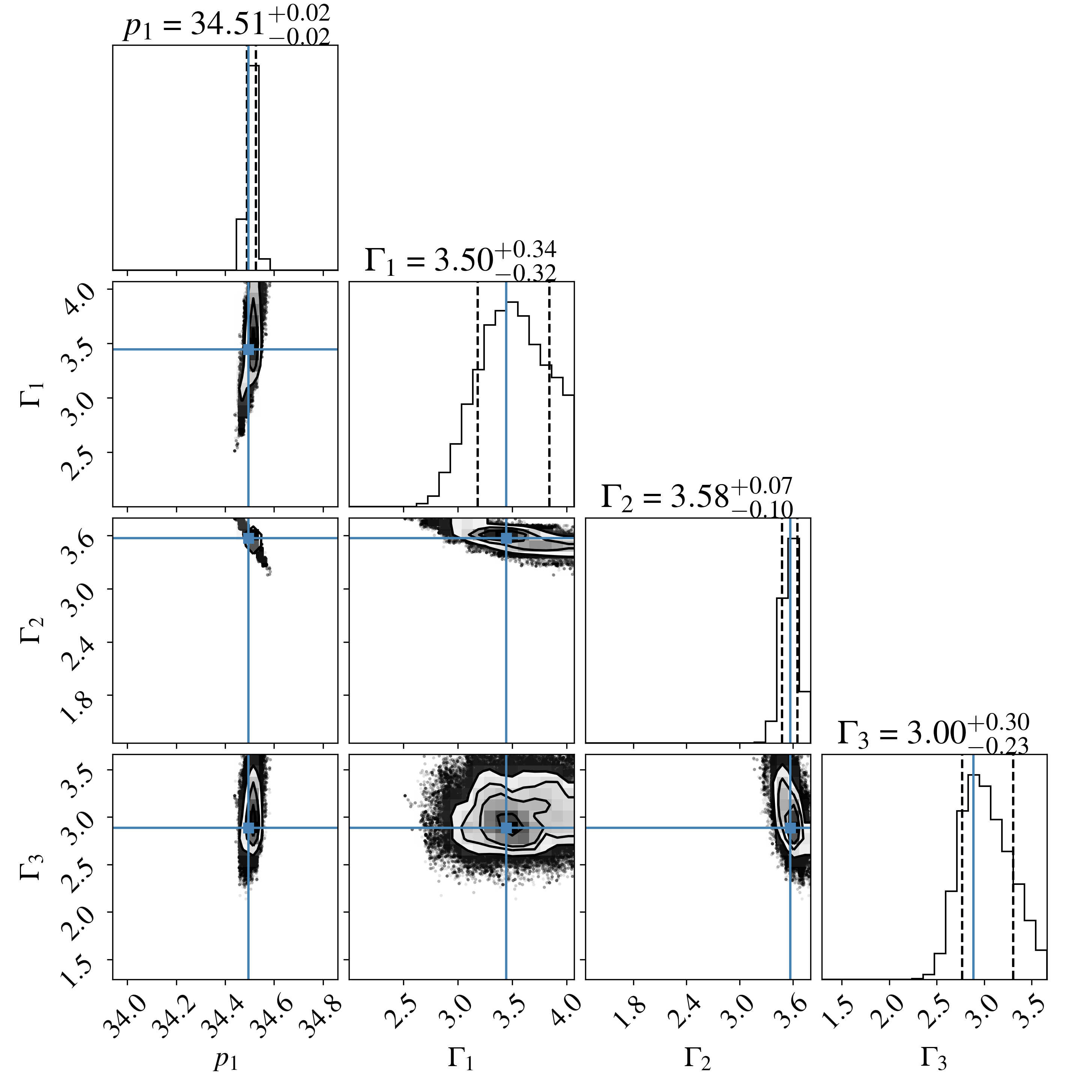}
\\~\\
\includegraphics[width=1.0\linewidth]{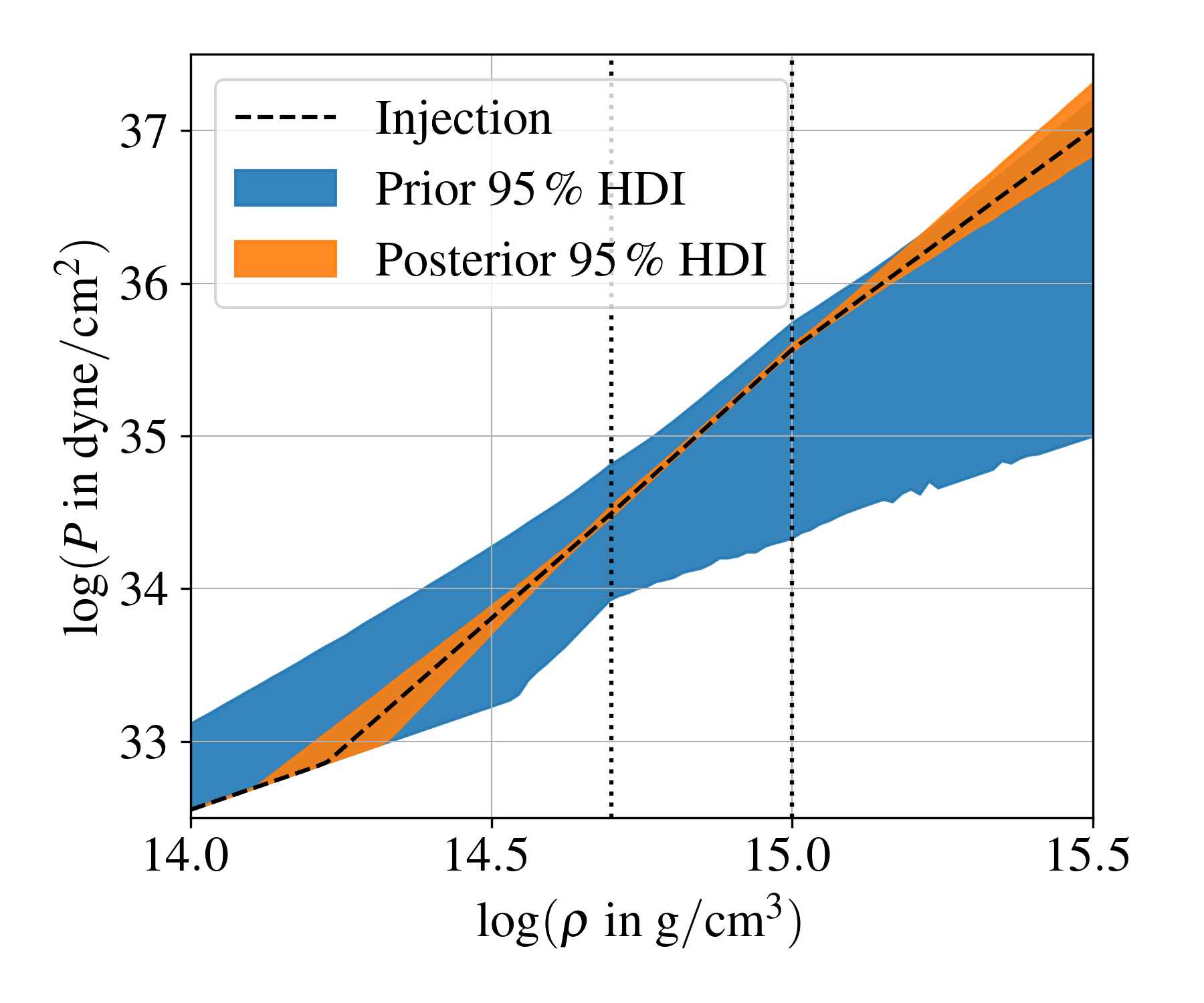}
\end{minipage}

\end{minipage}

\caption{
Results when providing the neutron star ``data'' using the slow rotation model used during the MCMC, not the more realistic one used in the main text. We use the same number of observed (simulated) neutron stars, but consider more high mass stars. The EOS is given by $\theta^{\rm MPA1}$. In the left panels we assume a relative error of $3\,\%$, in the right panels an error of $1\,\%$. Results for the MCMC sampling are shown in the top panels, the one for prior and posterior sampling in the bottom panels.
}
\label{fig_app1}
\end{figure}

\end{widetext}

To demonstrate that the small bias in the MPA1 and SLy parameter reconstruction in Sec.~\ref{app_MPA1} and Sec.~\ref{app_SLy} is really due to the imprecision of the model not fully taking into account rotation, we here apply it to data produced by the model itself, as described in Sec.~\ref{disc0}. We assume again the MPA1 EOS as injection, but change the observational data to include also heavier stars. Since all observables here have been produced with the model, we expect an even better reconstruction of the injected parameters. In Fig.~\ref{fig_app1} we show the results for the MCMC sampling and the prior and posterior sampling for the EOS. The results here also demonstrate that the $\Gamma_3$ parameter can also be recovered for the MPA1 EOS, if heavy enough stars have been observed.

\subsection{Results for Larger Measurement Errors}
\label{appendix_5}

\begin{widetext}

\begin{figure}

\begin{minipage}{1.0\linewidth}

\begin{minipage}{0.47\linewidth}
\includegraphics[width=1.0\linewidth]{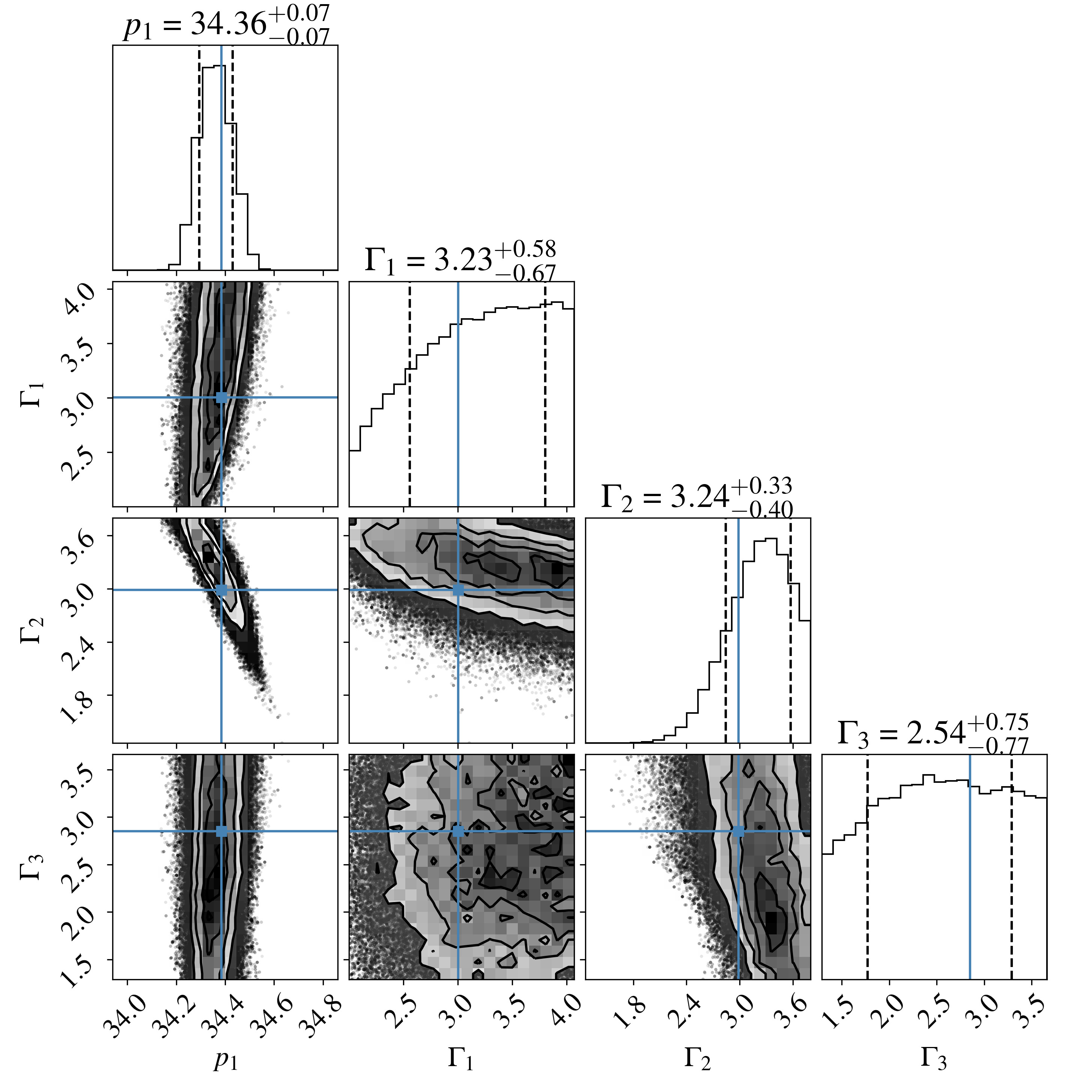}
\\~\\
\includegraphics[width=1.0\linewidth]{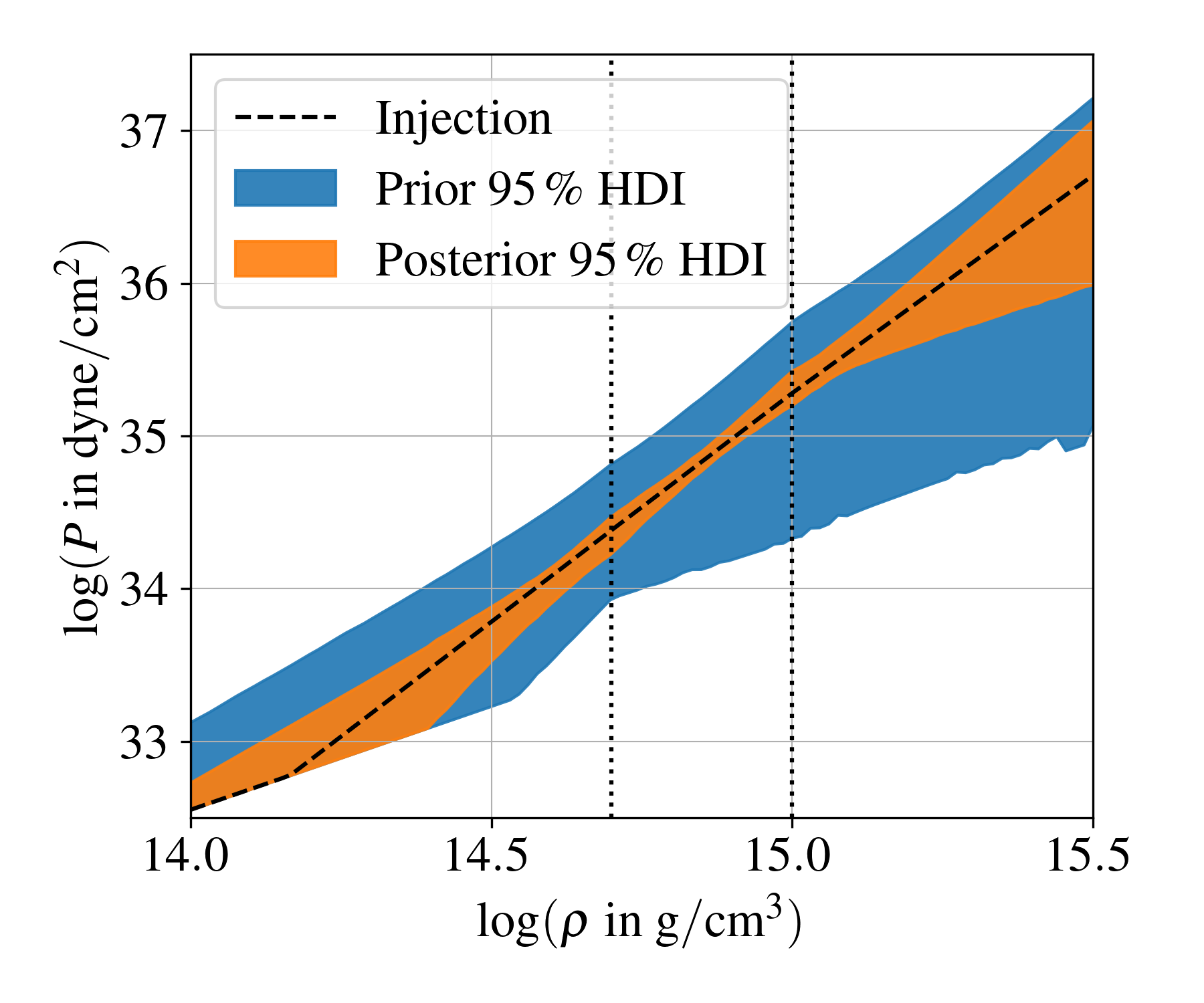}
\end{minipage}
\hfill
\begin{minipage}{0.47\linewidth}
\includegraphics[width=1.0\linewidth]{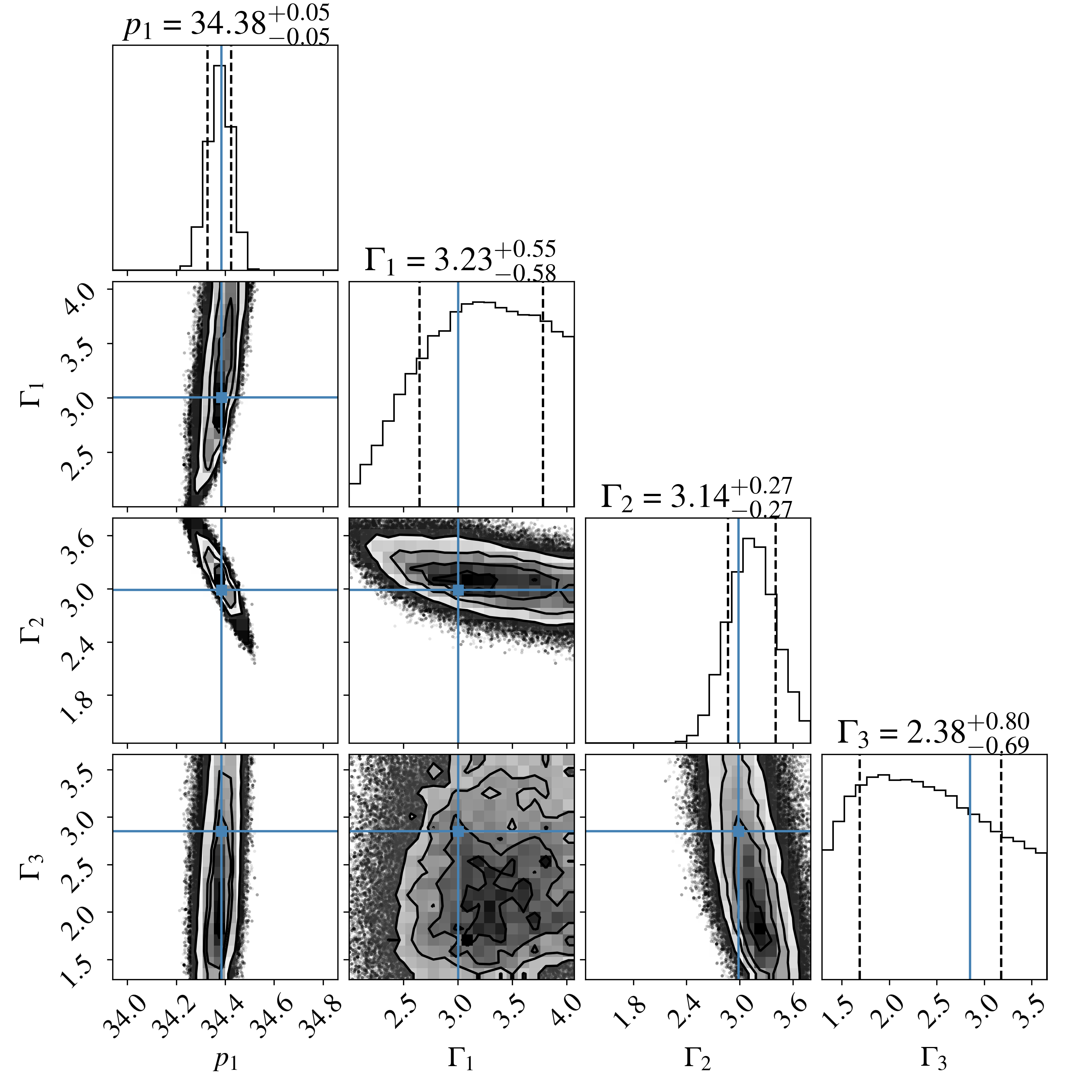}
\\~\\
\includegraphics[width=1.0\linewidth]{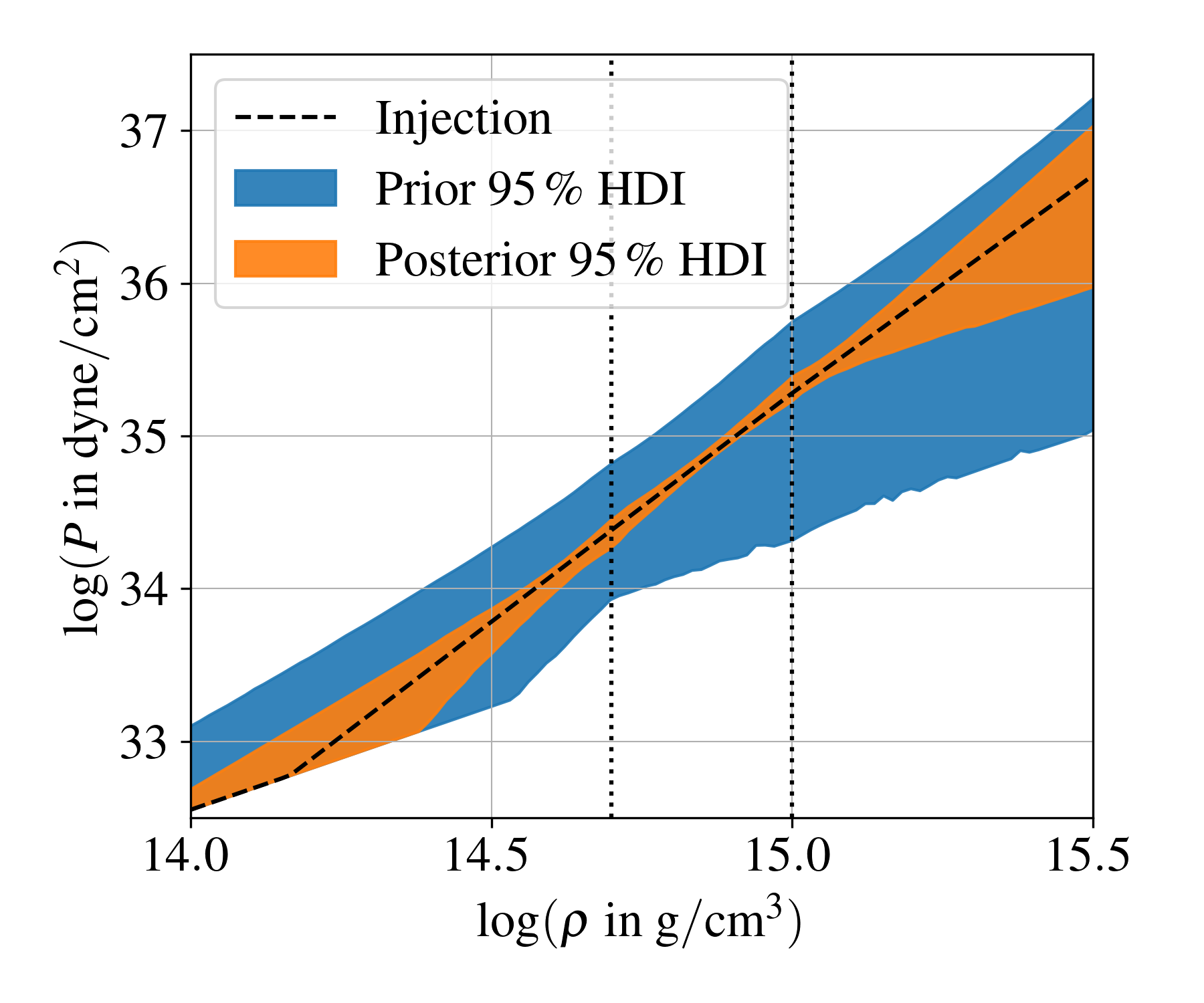}
\end{minipage}

\end{minipage}

\caption{
Results when using the same data as in the main text for the SLy EOS, but considering larger relative errors. In the left panels we assume a relative error of $10\,\%$, in the right panel of $5\,\%$. Results for the MCMC sampling are shown in the top panels, the one for prior and posterior sampling in the bottom panels.
}
\label{fig_app2}
\end{figure}

\end{widetext}

Since the examples in the main text assumed relative errors of $3\,\%$, we here show results for the SLy EOS where all observables are known with $5\,\%$ and $10\,\%$ relative errors in Fig.~\ref{fig_app2}. The data is the same ``realistic'' one as shown in Table~\ref{table_2}. As expected the bounds for the parameters become larger, $\Gamma_1$  and $\Gamma_3$ are basically not constrained. However, note that the bias described in Sec.~\ref{disc0} is smaller now, all parameters fall within the $[0.16, 0.84]$ quantils. Comparing the posterior samples with the ones of the prior in Fig.~\ref{fig_app2} shows that the data is still very informative, in particular in the intermediate density region.

\clearpage

\bibliography{literature}

\end{document}